\documentclass[a4paper,11pt]{article}
\usepackage{pos}

\usepackage{lipsum}						
\usepackage{comment}					
\usepackage{hyperref}					
\usepackage{framed}						
\usepackage{setspace}					
\usepackage{environ}                    

\usepackage{amsmath}					
\usepackage{amsfonts}					
\usepackage{mathtools}					
\usepackage{eqparbox}

\usepackage{physics}					
\usepackage{siunitx}					
\usepackage{slashed}                    
\usepackage{bbold}                      

\usepackage{float}                      
\usepackage{graphicx}					
\usepackage[labelfont=bf]{caption}      
\usepackage[width=.95\textwidth]{subcaption}    

\usepackage{booktabs} 
\usepackage{multirow} 

\newcommand{\upp}[1]{_{}^{#1}} 
\newcommand{\ind}[2]{^{#1}_{#2}} 

\makeatletter
\def\hlinewd#1{%
\noalign{\ifnum0=`}\fi\hrule \@height #1 \futurelet
\reserved@a\@xhline}
\makeatother
\title{Parton cascades at DLA: the role of the evolution variable}
\ShortTitle{Parton cascades at DLA: the role of the evolution variable}

\author[a]{Carlota Andrés}
\author[b,c]{Liliana Apolinário}
\author[d]{Néstor Armesto}
\author*[b,c]{André Cordeiro}
\author[d]{Fabio Dominguez}
\author[b,c]{José Guilherme Milhano}

\affiliation[a]{CPHT, CNRS, École polytechnique, Institut Polytechnique de Paris \\ 91120 Palaiseau, France}

\affiliation[b]{Laboratório de Instrumentação e Física Experimental de Partículas (LIP)  \\ Avenida Professor Gama Pinto, 2, 1649-003 Lisboa, Portugal}

\affiliation[c]{Departamento de Física,  Instituto Superior Técnico, Universidade de Lisboa \\ Avenida Rovisco Pais 1, 1049-001 Lisboa, Portugal}

\affiliation[d]{Instituto Galego de Física de Altas Enerxías (IGFAE), Universidade de Santiago de Compostela \\ Santiago de Compostela 15782, Spain}

\emailAdd{carlota.andres-casas@polytechnique.edu}
\emailAdd{liliana@lip.pt}
\emailAdd{nestor.armesto@usc.es}
\emailAdd{andre.cordeiro@tecnico.ulisboa.pt}
\emailAdd{fabio.dominguez@usc.es}
\emailAdd{gmilhano@lip.pt}

\abstract{
While experimental studies on jet quenching have achieved a large sophistication, the theoretical description of this phenomenon still misses some important points. One of them is the interplay of vacuum-like emissions, usually formulated in momentum space, with the medium induced ones that demand an interplay with a space-time picture of the medium and thus must be formulated in position space. A unified description of both vacuum and medium-induced emissions is lacking. In this work, we compute the tree-level probability of a double gluon emission in vacuum, and identify the enhanced phase-space regions for each diagram, corresponding to different configurations of the parton cascade. This calculation provides a parametric form for the formation times associated with each diagram, highlighting the equivalence of various ordering variables at double logarithmic accuracy. This equivalence is further explored by building a toy Monte-Carlo parton shower ordered in formation time, virtuality, transverse momentum, and angle. Aiming at a link with jet substructure, we compute the Lund Plane distributions and trajectories for each ordering prescription. We also compute the distributions in number of splittings and final partons, with the goal of clarifying the differences to be expected from the different ordering variables and the vetoes that must be implemented at Monte Carlo level to conserve energy-momentum, which turn out to have a sizeable influence on the shower's evolution.
}

\FullConference{%
    HardProbes2023\\
    26-31 March 2023 \\
    Aschaffenburg, Germany\\
}

\begin{document}

\maketitle

\section{Introduction}
\vspace{-0.25cm}
Ultra-relativistic heavy ion colliders, such as the Large Hadron Collider (LHC) and the Relativistic Heavy Ion Collider (RHIC) have unlocked the study of the Quark-Gluon Plasma (QGP), a hot and dense nuclear medium characterised by color deconfinement.
%
%
Studies of this state of matter are made possible by the analysis of high-energy jets produced in the collision. Compared to proton-proton collisions, their energy and substructure are modified through interactions with the medium(see e.g.~\cite{Cunqueiro:2021wls,Apolinario:2022vzg}). Due to the wide range of energy scales present within a jet, jet quenching observables may enable a space-time tomography of the QGP (e.g, ~\cite{Andres:2016iys,Apolinario:2017sob,Apolinario:2020uvt,He:2022evt}). A theoretical description of jet quenching effects requires a full accounting of the interplay between vacuum-like and medium-induced emissions~\cite{Caucal:2018dla} (see also \cite{Andres:2023hp} for a review), complicated by the fact that the former are formulated in momentum space by sampling a resummed no-emission probability, while the latter require an interface with a space-time picture of an expanding medium.
Aiming at such a description, this work gives an account of the vacuum evolution of parton cascades within the double logarithmic approximation (DLA), relying upon the ordering of successive parton emissions along some kinematic variable. The ambiguity inherent to this choice of ordering variable is reflected in the shower kinematics \cite{Nagy:2014nqa} and can be understood as characterising the uncertainty of this approximation. 
%
%
In this work we study this uncertainty by building three toy Monte Carlo parton showers ordered in virtual mass, formation time, and opening angle~\cite{Sjostrand:2006za, Dokshitzer:1991wu,Corcella:2000bw}. 
We explore the quark branch of the resulting cascades, comparing their kinematic distributions, Lund Plane~\cite{Dreyer:2018nbf} densities and trajectories, and evaluating the varying impact of a simplified jet quenching model.
\vspace{-0.30cm}

\section{Building differently ordered cascades}


%
The scale evolution of a QCD cascade is generated by sampling the \textit{no-emission probability} between a previous scale $s_{\rm prev}$ and the next scale $s$ , given by the Sudakov form factor,
\begin{align}
    \Delta(s_{\rm prev}, s_{}) = 
	\exp \left\{
	-\frac{\alpha_{s} C_{R} }{\pi} 
	\int^{s_{\rm prev}}_{s} \frac{\textrm{d} \mu}{\mu}
	\int^{}_{z(1-z) > z_{\rm cut}(\mu)} \frac{ \textrm{d} z }{z}
	\right\}
    \,,
\label{eq:no-emission-probability}
\end{align}
%
%
\noindent obtained by resumming the differential rate for resolvable emissions calculable in perturbative QCD. This requires integrating over all possible emission scales $\mu$ and splitting fractions $z$ and taking the leading logarithmic contribution for the splitting function, $\hat{P}(z) \simeq \frac{2 C_{R}}{z}$, where $R$ stands for the color representation of the emitter.
%
%
The integration range in \eqref{eq:no-emission-probability} is due to the resolution criterion for the splittings, corresponding to a transverse momentum cutoff. Explicitly, for some splitting $p \longrightarrow k_{1} + k_{2}$, this reads
\begin{align}
    |\boldsymbol{\ell}|\upp{2} 
    = z(1-z) p\upp{2} - (1-z) k^{2}_{1} - z k^{2}_{2}
    > k\ind{2}{\rm had}
    \,,
\label{eq:resolution-criterion}
\end{align}
\noindent where $\boldsymbol{\ell}$ stands for the relative transverse momentum between the daughter particles, and $z$ for the light cone momentum fraction\footnote{%
Another valid choice would be to take $z$ as the energy fraction of the emitted particle in some specific frame. 
These choices introduced another source of ambiguity when building a parton shower, and are not changed for the purposes of this work.
} of one of the daughters,
\begin{align}
    z &= k^{+}_{1} / p^{+} \,, 
    \,\,\, \textrm{ with } \,\,\,
    \boldsymbol{\ell} = (1-z) \boldsymbol{k}_{1} - z \boldsymbol{k}_{2} \,,
    \,\,\, \textrm{ and } \,\,\,
    \boldsymbol{p} = \boldsymbol{k}_{1} + \boldsymbol{k}_{2}
    \,.
\label{eq:splitting-kinematics}
\end{align}

%
The scale $s$ can now be assigned to a chosen kinematic variable, and the momenta of the emitted particles extracted. In this work, we define three such variables, namely the inverse light-cone formation time $t^{-1}_{\rm f}$, the virtual mass squared of the emitter $\Tilde{m}^{2}_{}$, and the light-cone angle between the daughters $\Tilde{\theta}^{2}$,
\begin{align}
	t^{-1}_{\rm f} = \frac{ |\boldsymbol{\ell}|\upp{2} }{ 2\,p\ind{+}{} z(1-z) } 
	\,, \quad
	\Tilde{m}^{2} = 
    \frac{ |\boldsymbol{\ell}|\upp{2} }{ z(1-z) } 
	\,,  \quad
	\Tilde{\theta}^{2}
	= 
    \left| \frac{\boldsymbol{k}^{}_{1}}{k\ind{+}{1}} - \frac{\boldsymbol{k}^{}_{2}}{k\ind{+}{2}} \right|\upp{2}
	= \frac{ |\boldsymbol{\ell}|\upp{2} }{ (p\ind{+}{})\upp{2} [z(1-z)]\upp{2} }
	\,.
\end{align}

According to each of these definitions one can write the resolution condition \eqref{eq:resolution-criterion} as a function of the splitting scale. As an example, in a formation time ordered shower one would write
\begin{align}
    |\boldsymbol{\ell}|^{2} > k^{2}_{\rm had} 
    \Longleftrightarrow z(1-z) > k^{2}_{\rm had} / (2 p^{+} t^{-1}_{\rm f}) \, ,
\label{eq:splitting-phase-space}
\end{align}
\noindent yielding the integration range in equation~\eqref{eq:no-emission-probability}, and providing an analytical expression for the available phase-space for emissions, expected to close as the radiating parton approaches the hadronisation scale.
The method for generating a parton splitting is straightforward. One begins by sampling the ordering variable from the no-emission probability \eqref{eq:no-emission-probability}, and sampling the light-cone momentum fraction from the splitting function, in the range given by the resolution condition $|\boldsymbol{\ell}|^2 > k^{2}_{\rm had}$, such that the ordering variable sets the regulator for the $z$ divergence. From these quantities one can obtain the relative transverse momentum $|\boldsymbol{\ell}|$ and extract the kinematics of the daughters by inverting the relationships in \eqref{eq:splitting-kinematics}.

This algorithm was implemented for the three ordering variables defined above, with the initial condition set by the scale of the hard scattering, written $t^{-1}_{\rm f} < p^{+}_{\rm jet}$, to express the scale separation between parton production and parton radiation. Further, as a consistency check, it was required that a generated splitting obeyed the condition $\Tilde{\theta}^{2} < 8 \Longrightarrow t^{-1}_{\rm f} < p^{+}$, also serving to avoid unreasonably wide splittings. This condition was implemented as a veto, such that any sampled $(s_{\rm trial}, z_{\rm trial})$ pair not obeying $\Tilde{\theta}^{2} < 8 $ was ignored, and the parton evolution continued to some lower scale $s < s_{\rm trial}$.

In the remainder of this work we describe some preliminary results for our parton cascades sampled according the three ordering prescriptions, for an initiating quark with light cone momentum $p^{+}_{\rm jet} / \sqrt{2} = 1000 \,\textrm{GeV}$, and an hadronization cutoff of $k_{\rm had} = 1 \,\textrm{GeV}$. 

\section{Comparing ordering prescriptions}

\begin{figure}[h]
\centering
\begin{subfigure}{.5\textwidth}
    \centering
    \includegraphics[width=\textwidth]{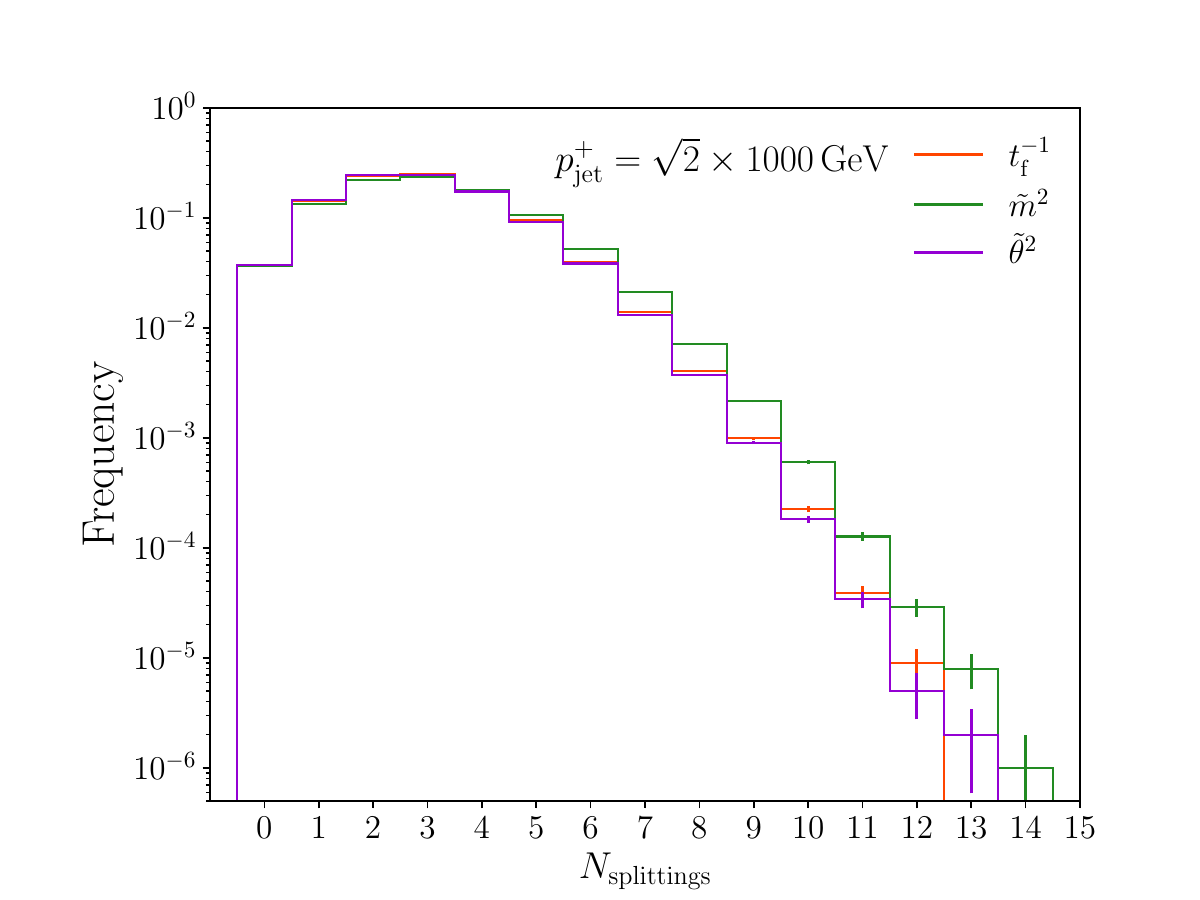}
    \label{subfig:nsplits-dist}
\end{subfigure}%
\begin{subfigure}{.5\textwidth}
    \centering
    \includegraphics[width=\textwidth]{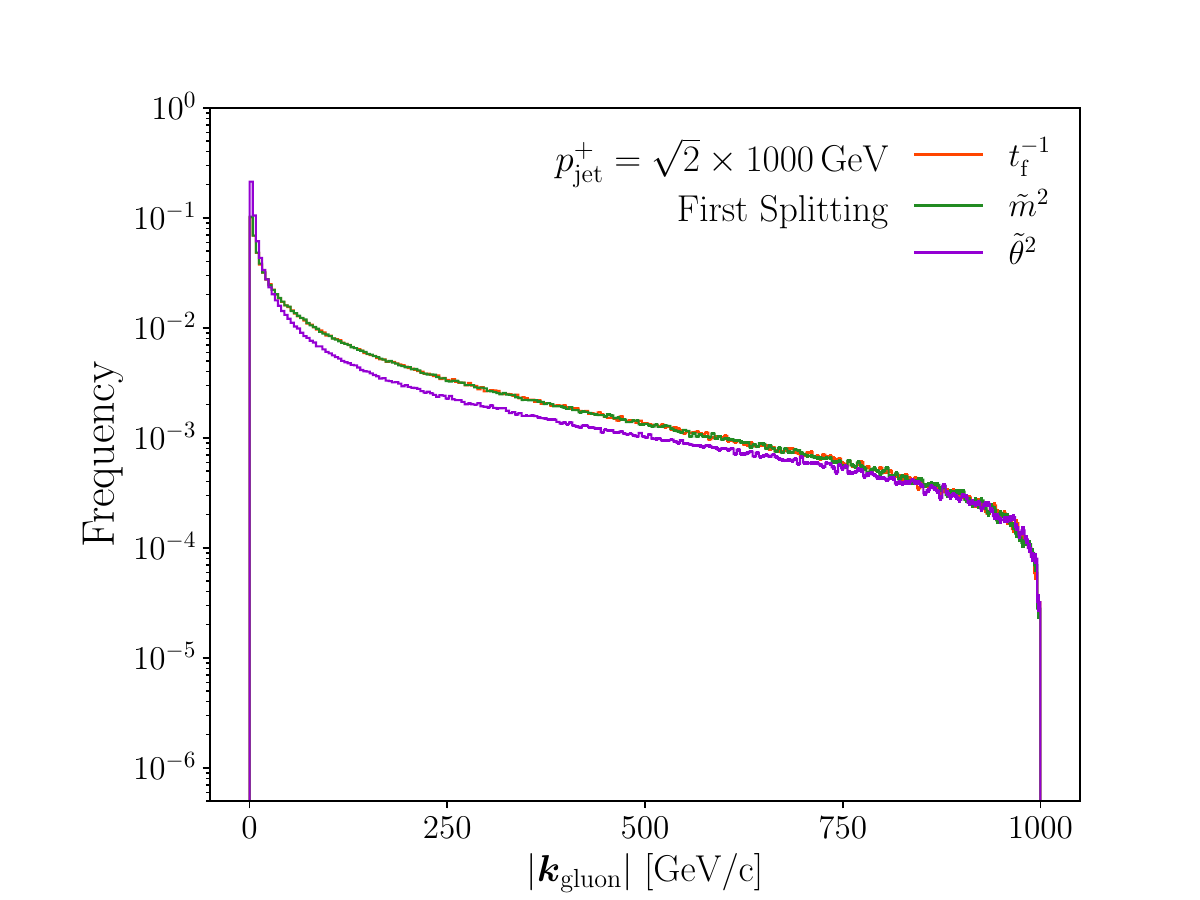}
    \label{subfig:ptgluon-dist}
\end{subfigure}
\caption{Distributions characterising the variation between parton cascades sampled according to different ordering prescriptions, namely formation time (orange), virtual mass (green), and light-cone angle (purple). \textbf{(Left)} Number of splittings along the quark branch. \textbf{(Right)} Transverse momentum of the first emitted gluon.}
\label{fig:1d-distributions}
\vspace{-0.3cm}
\end{figure}

The similarities between ordering prescriptions can be illustrated by considering some shower variables such as the number of gluon emissions along the quark branch of the cascade and the transverse momentum of the first emitted gluon, whose distributions are shown in figure~\ref{fig:1d-distributions}.

The number of quark splittings, in figure \ref{fig:1d-distributions} (left), characterises how the ordering prescriptions fill the available phase-space at different rates, with virtuality ordered showers (in green) resulting in longer cascades than formation time (in orange) and angular ordered showers (in purple) on average.
Despite these differences, the transverse momentum of the first emitted gluon (with respect to its parent) is distributed similarly across ordering variables, displaying the familiar logarithmic enhancement in figure~\ref{fig:1d-distributions} (right). Finding a similar behaviour for the three ordering variables with respect to these distributions, we turn to the characterisation of the shower evolution as a function of the splitting kinematics.




\section{Lund Plane densities and trajectories}



\begin{figure}[h]
\centering
\begin{subfigure}{.5\textwidth}
    \centering
    \includegraphics[width=\textwidth]{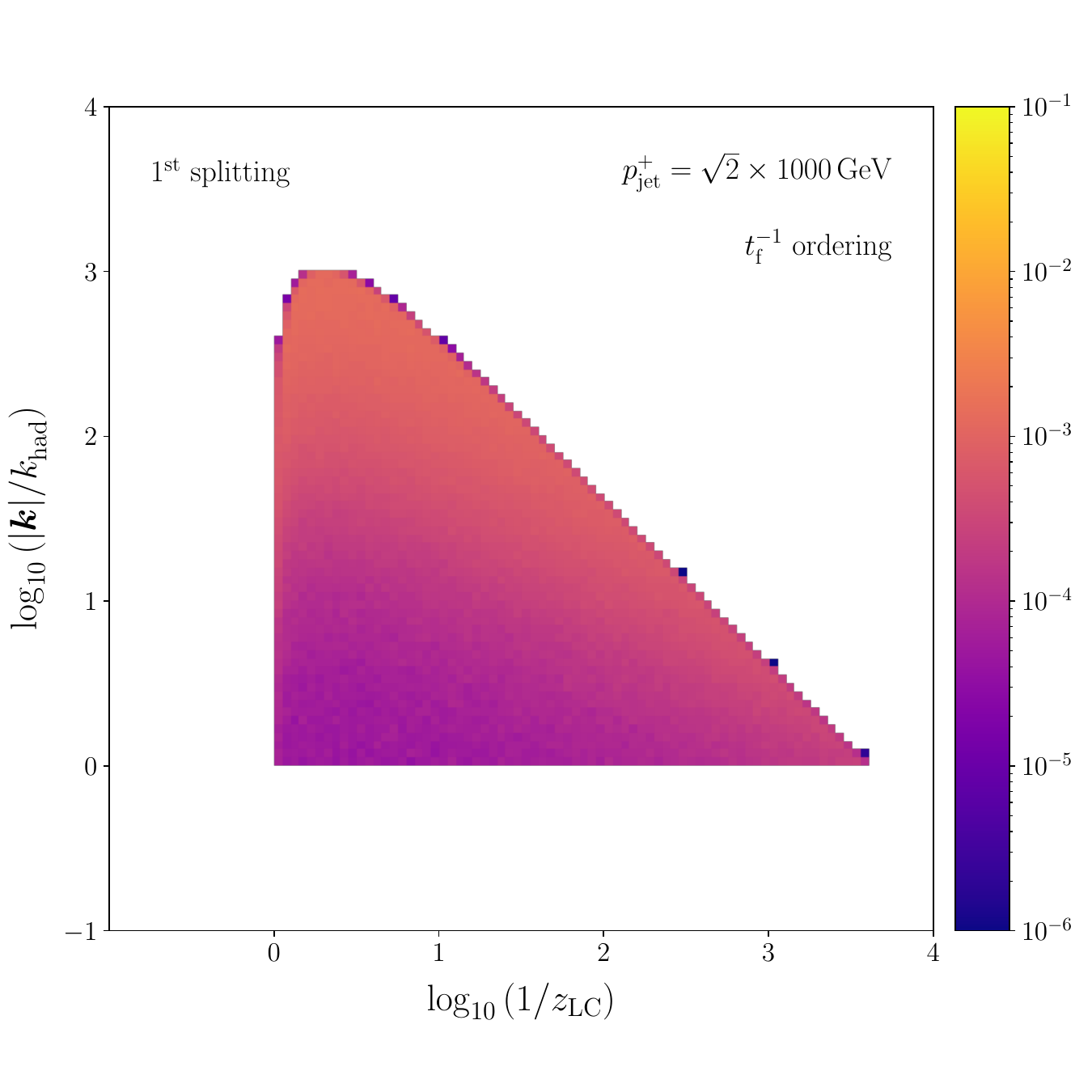}
    \label{subfig:lund-plane}
\end{subfigure}%
\begin{subfigure}{.475\textwidth}
    \centering
    \includegraphics[width=\textwidth]{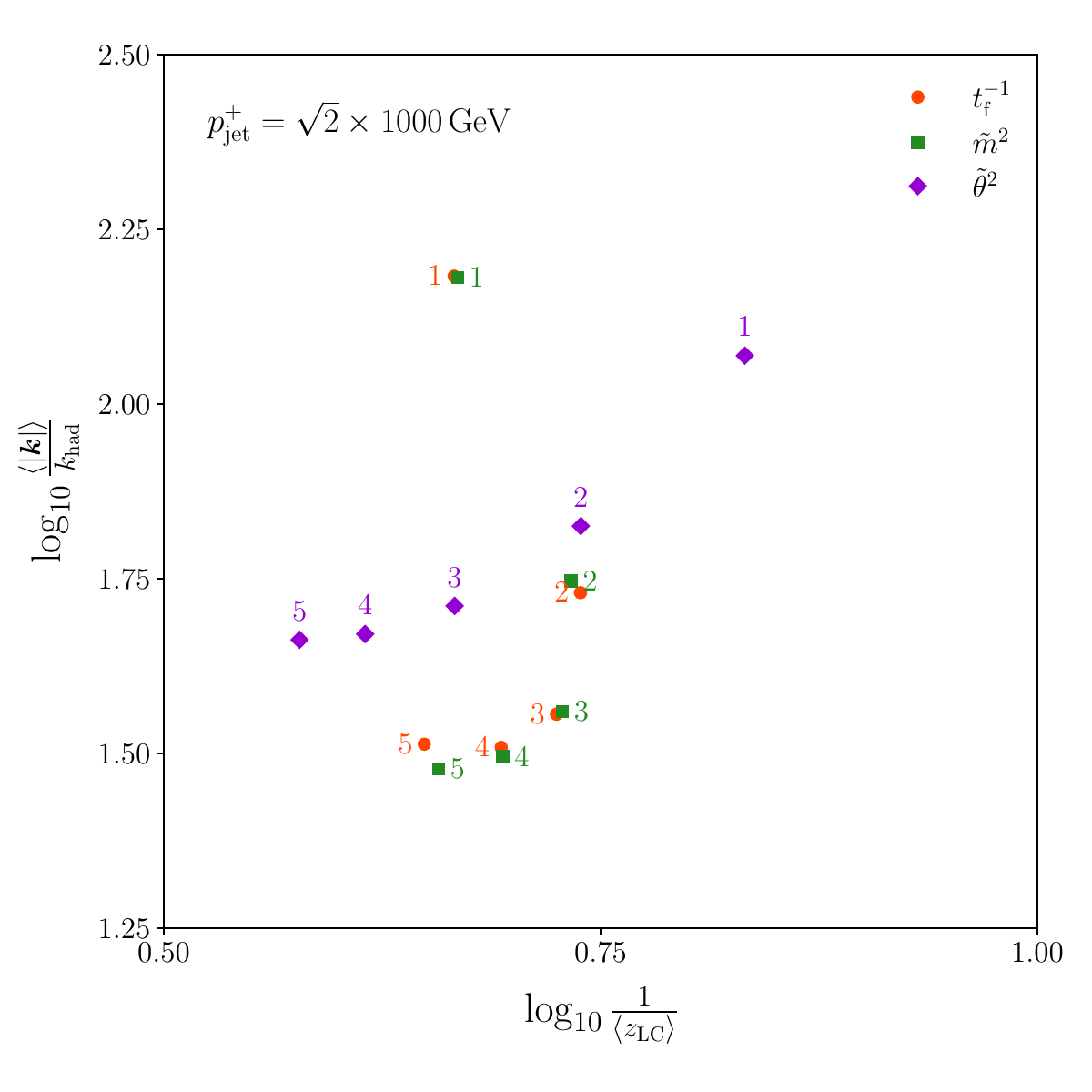}
    \label{subfig:lund-traj}
\end{subfigure}
\vspace{-0.5cm}
\caption{\textbf{(Left)} First splitting Lund distribution for cascades ordered in formation time. \textbf{(Right)} Lund trajectories for cascades ordered in formation time (orange), virtual mass (green), and angle (purple).}
\vspace{-0.3cm}
\label{fig:lund-plots}
\end{figure}

Lund diagrams \cite{Dreyer:2018nbf} allow for a representation of of parton emissions as points in a plane with coordinates $\big( \log(|\boldsymbol{k}| / k_{\rm had}), \log(1/z) \big)$.
%
%
%
As such, figure~\ref{fig:lund-plots} (left) depicts the Lund plane density for the first emission of cascades ordered in formation time, generally favoring large transverse momenta and small light cone fractions. A comparison between ordering prescriptions is made simpler if one considers the ``trajectories'' of the quark splittings in this Lund plane that is, the mean values of the $|\boldsymbol{k}|$ and $z$ distributions depicted for different splittings in a Lund diagram. This is shown in figure \ref{fig:lund-plots} (right), for the first five quark splittings and all three ordering prescriptions. Here we find an evolution towards increasing values of the light-cone fraction (relative to the quark), consistent with the closing of the phase-space predicted in equation~\eqref{eq:splitting-phase-space}.
%
%
It is also worth noting that showers ordered in formation time (in orange) and virtual mass (in green) follow similar trajectories, with the difference arising due to the relative factor of $p^{+}$ in the variable definition, while the angular ordered shower (in purple) evolves along a different path, roughly corresponding to increasing angle.

%
%

In general, this means differently ordered cascades can differ significantly in their kinematics, resulting in a different distribution of formation times, $t^{-1}_{\rm f}$. In order to better explore these differences, which may play a determining role in coupling the shower evolution with a finite medium, the next section explores a simplified quenching model, and its impact on differently ordered cascades.



\section{A simple quenching model}


A quenching model can be built by simplifying the parton-medium interactions into a single condition, stating that a shower is suppressed if, in any of its splittings, the relative transverse distance acquired by the daughter partons is above the saturation scale of a static medium \cite{Mehtar-Tani:2012mfa}, taken to be $Q^{-1}_{\rm sat} \sim (\hat{q} L)^{-1/2}$, where the transport coefficient $\hat{q}$ depends on the medium density and $L$ stands for the medium length. This can be translated into the following quenching probability,
\vspace{-0.3cm}
\begin{align}
    \mathcal{P}_{\rm quenching} 
    = \Theta( L - t_{\rm f} ) \times 
    \Theta\left( {\rm d}_{\rm s} - \left(\hat{q} (L - t_{\rm f})\right)^{-1/2} \right)
    \textrm{ with }
    {\rm d}_{\rm s} = \sqrt{\frac{t_{\rm f}}{z \,p^{+}}}
    \,,
\label{eq:quenching-model}
\end{align}
\noindent where the medium length $L$ was shifted to account for a finite formation time. Further, only splittings produced inside the medium length, $L > t_{\rm f}$, can be suppressed by medium effects.


The simplified quenching model was applied to the cascades sampled according to all three ordering prescriptions. This was done in two different modes; either checking the conditions in in equation~\eqref{eq:quenching-model} only against the first emission in the cascade, or against the full quark branch. The percentage of quenched events are displayed in tables~\ref{subtab:delayed-brick-1st-splitting} for the former case and \ref{subtab:delayed-brick-fullshower} for the latter. In these tables, two different values for $L$ and three different values for $\hat{q}$ were considered.
%
%
While we find significant differences between the ordering variables when the model is applied exclusively to the first splittings, these are seen to disappear when the full quark branch is taken into account.


\begin{table}[h]
\begin{subtable}{.5\linewidth}
\centering
\begin{tabular}{|c||c|c|c|}
    \hline
    $L$ [fm]         & 
    \multicolumn{2}{c|}{4} & 6 \\    
    \hline
    $\hat{q}$ [${\rm GeV}^2$ / ${\rm fm}$]   & 
    2 & 5 & 5 \\
    \hline
    $t^{-1}_{\rm f}$	 & 1.1 \% 	 & 3.1 \%  	 & 5.9 \%  \\
    $\tilde{m}^{2}_{}$	 & 1.1 \% 	 & 3.1 \% 	 & 5.9 \%  \\
    $\tilde{\theta}^{2}_{}$	 & 4.0 \% 	 & 9.1 \% 	 & 15.6 \%  \\
    \hline
\end{tabular}
\subcaption{Applied only to the first emission.}
\label{subtab:delayed-brick-1st-splitting}
\end{subtable}
\hfill
\begin{subtable}{.5\linewidth}
\centering
\begin{tabular}{|c||c|c|c|}
    \hline
    $L$ [fm]         & 
    \multicolumn{2}{c|}{4} & 6 \\
    \hline
    $\hat{q}$ [${\rm GeV}^2$ / ${\rm fm}$]   & 
    2 & 5 & 5 \\
    \hline
    $t^{-1}_{\rm f}$	 & 4.6 \% 	 & 11.5 \% 	 & 22.0 \%  \\
    $\tilde{m}^{2}_{}$	 & 4.9 \% 	 & 12.4 \% 	 & 23.5 \%  \\
    $\tilde{\theta}^{2}_{}$	 & 4.6 \% 	 & 11.5 \% 	 & 22.0 \%  \\
    \hline
\end{tabular}
\subcaption{Applied to the full quark branch.}
\label{subtab:delayed-brick-fullshower}
\end{subtable}
\caption{Percentage of quenched events when applying the simplfied model to differently ordered cascades.}
\label{tab:quench-probs}
\end{table}

These results indicate that while the integrated energy loss of a parton cascade is not sensitive to the ordering prescription at DLA accuracy, medium-induced effects (or modifications) over the first few splittings may be affected by this choice. Compounded by lack of medium evolution in this simplified model, these findings motivate the need for a theoretically consistent formalism to describe the interface between a developing jet and an evolving medium.

\vspace{-0.25cm}
\section{Summary}
In this work we developed a toy Monte Carlo for parton showers at double logarithmic accuracy. The setup allows to consistently change between different ordering variables that were chosen to be formation time, virtual mass and opening angle.
%
%
The Lund plane densities and trajectories for all three prescriptions were computed. Moreover, we applied a simplified quenching model inspired by coherence effects in a finite size medium of constant density and obtained the suppression probability for different quenching scenarios. While this quenching probability was found to be independent of the ordering prescription when the quenching model is applied to the full shower, there are significant differences when applying the model exclusively to the first emission, motivating the necessity of an interface between jet development and medium evolution.

\vspace{-0.25cm}
\section*{Acknowledgements}
This work has received funding by OE Portugal, Fundação para a Ciência e a Tecnologia (FCT), I.P., projects EXPL/FIS-PAR/0905/2021 and CERN/FIS-PAR/0032/2021; by European Research Council project ERC-2018-ADG-835105 YoctoLHC and has received funding from the European Union’s Horizon 2020 research and innovation programme under grant agreement No.824093. A.C. and L.A. were directly supported by FCT under contracts PRT/BD/154190/2022 and 2021.03209.CEECIND.
C.A. has received funding from the European Union’s Horizon 2020 research and innovation program under the Marie Skłodowska-Curie grant agreement No 893021 (JQ4LHC).
%
%
N.A. has received financial support from Xunta de Galicia (Centro singular de investigación de Galicia accreditation 2019-2022), by European Union ERDF, and by the Spanish Research State Agency under project PID2020-119632GB-I00.
\vspace{-0.3cm}
\bibliographystyle{JHEP}
\bibliography{bibliography.bib}

\end{document}